\documentclass{aa}  
\bibpunct{(}{)}{;}{a}{}{,} % to follow the A&A style
\usepackage{graphicx}
\usepackage{txfonts}
\usepackage{natbib}
\newcommand{\be}{\begin{equation}}
\newcommand{\ee}{\end{equation}}
\newcommand{\bea}{\begin{eqnarray}}
\newcommand{\eea}{\end{eqnarray}}

\begin{document} 

\authorrunning{C. E. Alissandrakis etal}
\title{Modeling the quiet Sun cell and network emission with ALMA
}
%\subtitle{Implications for solar atmospheric models}
\author{C. E. Alissandrakis$^1$, A. Nindos$^1$, T. S. Bastian$^2$, \and S. Patsourakos$^1$
}

\institute{Department of Physics, University of Ioannina, GR-45110 Ioannina, 
Greece\\
\email{calissan@cc.uoi.gr}
\and
National Radio Astronomy Observatory (NRAO), 520 Edgemont Road, Charlottesville, VA 22903, USA
}

\date{Received ...; accepted ...}

% \abstract{}{}{}{}{} 
% 5 {} token are mandatory
 
  \abstract
{ALMA observations of the Sun at mm-$\lambda$ offer a unique opportunity to investigate the temperature structure of the solar chromosphere. In this article we expand our previous work on modeling the chromospheric temperature of the quiet Sun, by including measurements of the brightness temperature in the network and cell interiors, from high resolution ALMA images at 3\,mm (Band 3) and 1.26\,mm (Band 6). We also examine the absolute calibration of ALMA full-disk images. We suggest that the brightness temperature at the center of the solar disk in Band 6 is $\sim440$\,K above the value recommended by \cite{2017SoPh..292...88W} and we give improved results for the electron temperature variation of the average quiet Sun with optical depth, as well as the derived spectrum at the center of the disk. We found that the electron temperature in the network is considerably lower than predicted by model F of \cite{1993ApJ...406..319F} and that of the cell interior considerably higher than predicted by model A. Depending upon the network/cell segregation scheme, the electron temperature difference between network and cell at $\tau=1$ (100\,GHz) is from $\sim$660 to $\sim$1550\,K, compared to $\sim$3280 K predicted by the models; similarly, the $T_e$ ratio is from $\sim$1.10, to 1.24, against $\sim$1.55 of the model prediction. We also found that the network/cell $T_e(\tau)$ curves diverge as $\tau$ decreases, indicating an increase of contrast with height and possibly a steeper temperature rise in the network than in the cell interior.
}


   \maketitle
%
%________________________________________________________________

\section{Introduction}\label{intro}

Our knowledge on the physical conditions of the upper solar atmosphere is based primarily on extreme ultraviolet (EUV) observations. Although the same atmospheric region emits in the radio range as well, older radio data suffered from low spatial resolution and absolute calibration problems, which limited their usefulness in modeling. 

The Bilderberg Continuum Atmosphere \citep[BCA,][]{1968SoPh....3....5G} was the first model to take into account
mm-wave observations; a comparison between BCA-predicted brightness temperatures and observations in the range of 0.0086 -- 15.8mm was presented in Fig.~7 of \cite{1968SoPh....3...36N}. This practice continued in subsequent models, such as the Harvard Smithsonian Reference Atmosphere \citep{1971SoPh...18..347G} and the VAL models \citep{1976ApJS...30....1V,1973ApJ...184..605V,1981ApJS...45..635V}, among others.

Starting with the model of \citet[hereafter VAL81]{1981ApJS...45..635V}, a multi-component approach was developed, aiming at describing the emission of fine atmospheric structures in the horizontal direction, such as the chromospheric network and cell interior (also known as intra-network). These models are not truly 3D, as radiative transfer in the horizontal direction is ignored, justified by the argument that the horizontal scale of the structures is much larger than the vertical. The few published measurements on the brightness of cell interiors and the network in the microwave and the mm-$\lambda$ range, reviewed by \cite{2011SoPh..273..309S}, indicate that the network/cell contrast increases with the wavelength. This increase is consistent with the computations of \cite{1983SoPh...85..237C}, based on the VAL81 model. 

With the advent of fast numerical computations, a number of sophisticated tools, such as the {\it Bifrost} radiative magnetohydrodynamics (rMHD) code \citep{2011A&A...531A.154G} and the {\it STockholm inversion Code} \citep[STic][]{2019A&A...623A..74D}) have been developed for solar atmospheric modeling. Such models have been employed in the analysis or mm-wavelength data by \cite{2004A&A...419..747L}, \cite{2007A&A...471..977W} and \cite{2020A&A...635A..71W}, among others. Nevertheless, the classic models still provide a clear and comprehensive picture of the solar atmosphere.

For a number of well-known reasons that we will not repeat here \citep[see][for a review]{2019AdSpR..63.1396L}, the Atacama Large Millimeter/submillimeter Array (ALMA) is the ideal instrument for probing the solar chromosphere in the mm-$\lambda$ range. In a previous article \citep[][hereafter Paper I]{2017A&A...605A..78A}, we inverted center-to-limb data for the average quiet Sun (QS) measured from full-disk (FD) ALMA images obtained during the commissioning period of December 2015 in Band 3 (100\,GHz) and Band 6 (239\,GHz), together with the observations of \cite{1993ApJ...415..364B} at 353\,GHz, to compute the variation of the electron temperature, $T_e$, as a function of the optical depth at 100\,GHz, $\tau_{100}$. We found that $T_e(\tau_{100})$ was close (5\% lower) to the prediction of model C of \cite{1993ApJ...406..319F}, hereafter FAL93.

In this work we expand our modeling to the cell interior and network elements, by including measurements from high-resolution (HR) ALMA images in Band 3 \citep[][hereafter Paper II]{2018A&A...619L...6N} and in Band 6, and we compare our results to multi-component models of the solar atmosphere. In Section~\ref{FD} we examine the normalization of FD images and we report improved results on the average QS and the height of the mm-$\lambda$ emission. In Section~\ref{HR} we report our results on the cell interior and the network. Finally, we summarize and discuss our results in Section~\ref{summary}.

\section{Full-disk ALMA images}\label{FD}
\subsection{Normalization of full-disk images}
ALMA employs a sophisticated system to scan the full solar disk with the four 12\,m dishes, described in detail by \cite{2017SoPh..292...88W}, providing full-disk (FD) images with a resolution of $\sim60$\arcsec\ in Band 3 and $\sim30$\arcsec\ in Band 6, over a field of view (FOV) of 2400\arcsec. Although this system gives high quality images, the absolute calibration is complicated due to the many instrumental and atmospheric parameters implicated; moreover, celestial calibrators cannot be used, as in the case of interferometric images. As a consequence, \cite{2017SoPh..292...88W} recommended that the FD images be normalized to particular values of brightness temperature at the center of the solar disk.

In Paper I, we did not apply the recommended normalization, but used the FD images as they were in the ALMA site. We also argued that non-zero emission beyond the limb was due to diffuse light, rather than due to the sky background; consequently zero sky background was assumed and the observed $T_b$ was corrected for diffuse light as explained in Paper I (Section 3 and Figure 4); this correction was small and only affected the region near the limb. The 353\,GHz data of \cite{1993ApJ...415..364B} were already normalized by setting the disk center $T_b$ to 5580 K. The disk center $T_b$ for Band 3 quoted in Paper I (7250\,K) is very close to the recommended value (7300\,K), whereas for Band 6 it is 280\,K above the recommended value (6180\,K, compared to 5900\, K), or higher by a factor of 1.047.

In order to combine data at different frequencies, we had reduced all data to a common reference frequency, $f_{ref}$, using the fact that both the free-free \citep{1970resp.book.....Z} and the H$^-$ \citep{1974ApJ...187..179S} absorption coefficients are proportional to $f^{-2}$. Hence a measurement at a frequency $f$ is remapped to
\be
T_b\left((f/f_{ref})^2\mu,f_{ref}\right)=T_b(\mu,f)
\label{reduce}
.\ee
where $\mu=\cos\theta$, $\theta$ being the heliocentric angle. We note that although the contribution of H$^-$ in the opacity is small ($\sim10$\% around $T_e=6000$\,K), it is not negligible.

\begin{figure}%[h]
\centering
\includegraphics[width=\hsize]{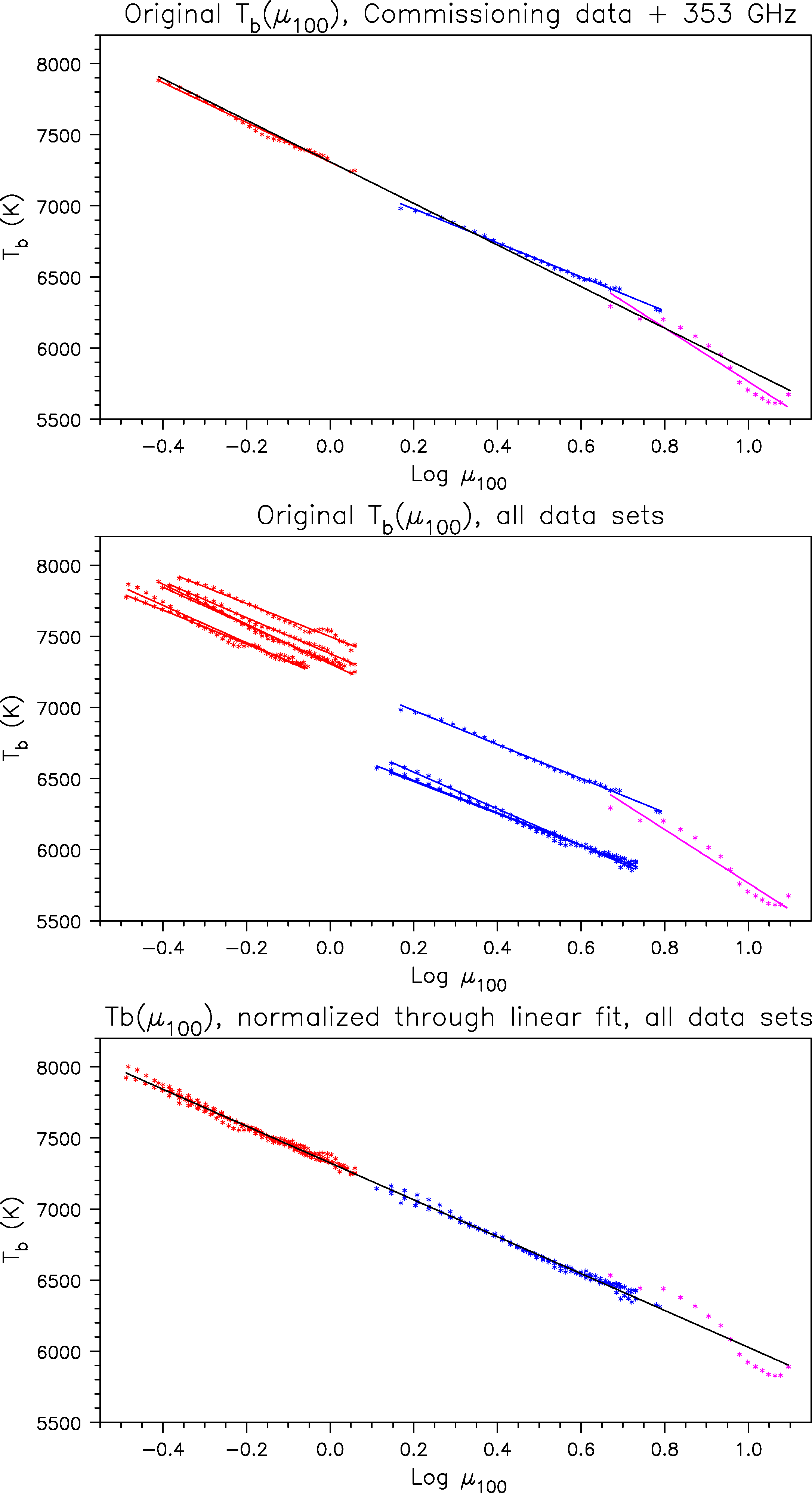}
\caption{Top: Original brightness as a function of reference $\mu$ ($\mu_{100}=\mu_{obs}\,(f_{obs}/100\mbox{GHz})^2$) for commissioning and 353\,GHz data only (red for band 3, blue for band 6, magenta for 353\,GHz). Middle: Original brightness for all data sets. Bottom: Normalized brightness for all data sets. Colored lines are linear fits to individual bands, black lines are fits to all data points.}
\label{norm}
\end{figure}

Following this procedure, $T_b$ plotted as a function of $\log \mu$, reduced to $f_{ref}=100$\,GHz, showed that the 3 data sets were consistent to one another (Fig.~\ref{norm} top; see also Fig. 5 of Paper I), thus making possible the inversion and the computation of $T_e(\tau)$. Note that in Paper I it was assumed that FD images were at the average frequency of each band; this was not correct, and the actual frequencies (107 instead of 100\,GHz for band 3, 248 instead of 233\,GHz for band 6) were used in Fig.~\ref{norm}. 

As we reported in Paper I, the empirical logarithmic dependence of the brightness temperature on $\mu$ implies a logarithmic dependence of the electron temperature on the optical depth,
\be
T_e(\tau)=a_1+a_2\ln\tau         \label{logform}
,\ee
for which the brightness temperature is, from the transfer equation:
\be
T_b(\mu)=a_1+a_2(\ln\mu-\gamma)
\label{tbmu}
,\ee
where $\gamma$ is the Euler constant. We note that in the plots presented in this article we preferred to use $\log\mu$ rather that $\ln\mu$ to make them more comprehensible.

The form of  (\ref{logform}) reflects the gradual temperature rise in the chromosphere; the actual $T_e(\tau)$ is expected to steepen at low $\tau$ as we reach the transition region and its slope to change again at large $\tau$, as we approach the temperature minimum.

For our April 12, 2018 observations, which will be presented in detail in a future publication, the FD images were normalized according to the recommended disk center values. When plotting $T_b(\mu)$, a very noticeable jump appeared between band 6 and the other bands. This is true for other data sets obtained during very quiet dates that we examined, listed in Table~\ref{Table02}, where N is the number of images used. Measurements from all of them are plotted together in the middle panel of Fig.~\ref{norm}; straight lines are linear fits to each individual band. The only explanation for this jump is that the normalization applied to band 6 is not correct. 

\begin{figure*}%[b]
\centering
\includegraphics[width=\textwidth]{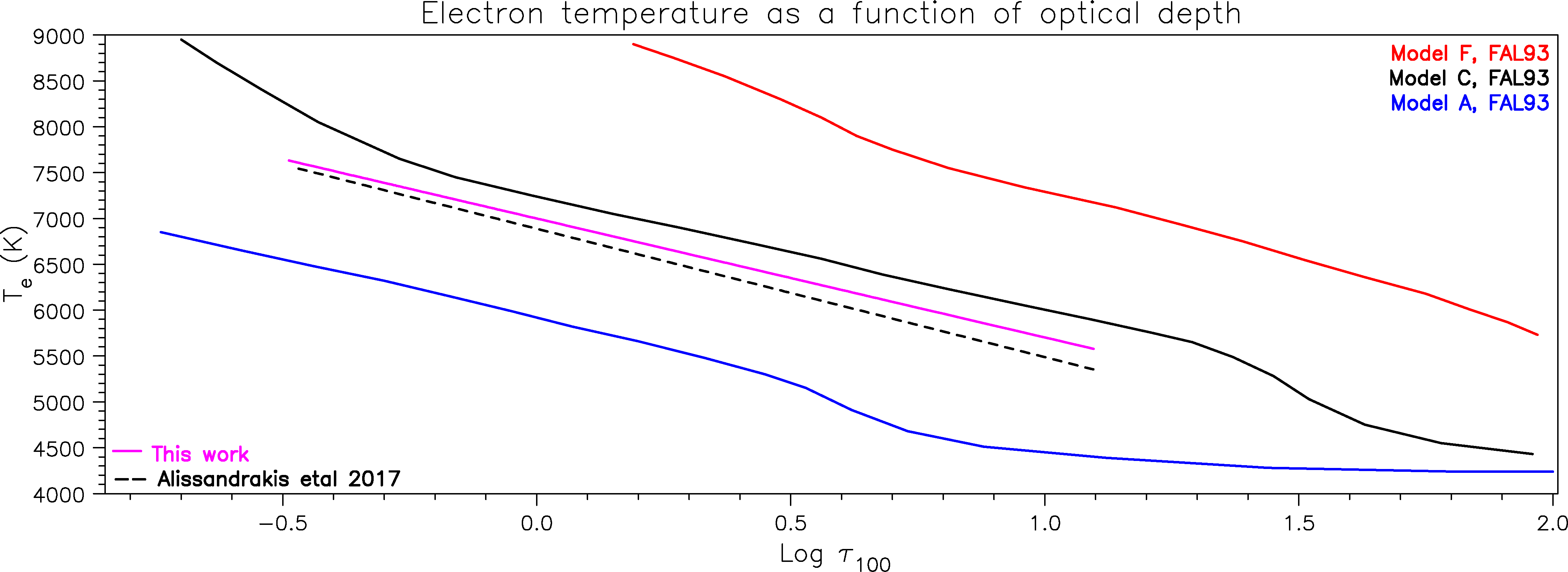}
\caption{The electron temperature as a function of $\tau_{100}$ from the current work (full magenta line), together with the FAL93 models A (blue), C (black) and F (red). The dashed black line shows the electron temperature reported in Paper I.}
\label{invert}
\end{figure*}

Normalizing to the commissioning disk center values gave a reasonable, but not good enough agreement among the 3 bands.
We found that the best way to normalize the data is through a least square fit of all sets to the same linear (or quadratic) function of $\log \mu$. Considering that Band 3 is more reliable than Band 6, both because our commissioning value is close to the recommended and because atmospheric conditions are easier to handle, we set the normalization factor for that at unity, leaving the factors for the other data sets to be determined by the fit. The result of the fit is very good, as shown in the bottom panel of Fig.~\ref{norm}, with a root mean square (rms) deviation of about 20\,K for the ALMA data. Here a linear fit function was used, as a quadratic fit made no difference.

\begin{table}[!h]
\begin{center}
\caption{Dates of ALMA FD data and normalization factors}
\label{Table02}
\begin{tabular}{lcccl}
\hline 
Date          & N&Freq&  Norm  & Origin of the data \\
          & &GHz&   &  \\
\hline
Dec 2015      &~5&107 & 1.0000 & Commissioning \\
Mar 16, 2017  &~9&107 & 0.9793 & Paper II \\
Apr 12, 2018  &10&107 & 0.9934 & This work \\
May  1, 2018  &14&~93 & 1.0169 & ALMA 2017.1.00870.S \\
Dec 20, 2018  &~5&~95 & 1.0188 & ALMA 2018.1.01763.S \\
Dec 20, 2018  &~4&105 & 1.0006 & ALMA 2018.1.01763.S \\
\hline 
Dec 2015      &~3&248 & 1.0085 & Commissioning \\      
Apr 12, 2018  &~7&232 & 1.0840 & This work \\
May  1, 2018  &11&232 & 1.0871 & ALMA 2017.1.00870.S \\
Dec 20, 2018  &~2&232 & 1.0864 & ALMA 2018.1.01763.S \\
\hline 
Jul 9-10, 1991  &~1&353 & 1.0383 & \cite{1993ApJ...415..364B} \\
\hline 
\end{tabular}
\end{center}
\end{table}

The derived normalization factors are given in Table~\ref{Table02}; for Band 6 they are $\sim8$\% which, for a $T_b$ value of 6000\,K translates to a difference of $\sim500$\,K. Additional information on this issue comes from comparing the average $T_b$ of ``feathered'' images (combining HR interferometric data with FD) to the one measured from FD images.  For our Band 6 observations of April 12, 2018 these values were not the same and, in order to correct for this difference, we had to multiply the FD values by a factor of 1.066. Noting that the calibration of the interferometric images, based on celestial sources, is more reliable than the FD calibration and that this correction is in the same direction as, and quite close to the value of 1.084 listed in Table~\ref{Table02}, this supports our conclusion that FD Band 6 values are underestimated.

\subsection{Inversion and spectrum}\label{inversion}
Inverting the selected data set we obtained new values of the inversion parameters in eq (\ref{tbmu}), $a_1$ and $a_2$, very close the values reported in Paper I; they are given in Table~\ref{Table03}, together with the values reported in Paper I. 
Statistical errors of the normalization fit are very small, of the order of $\delta a_1=2.6$\,K and $\delta a_2=1.4$\,K. A more realistic estimate is obtained from the dispersion of the inversion parameters when single-day observations are considered individually, which gives $\delta a_1=14$\,K and $\delta a_2=15$\,K.
 
\begin{table}[!h]
\begin{center}
\caption{Inversion parameters}
\label{Table03}
\begin{tabular}{lcc}
\hline 
Parameter& This work&Paper I  \\
\hline 
$a_1 (K)$ & 6999    &  6887 \\
$a_2 (K)$ & $-$563 & $-$608\\
\hline 
\end{tabular}
\end{center}
\end{table}

The inverted $T_e$ is plotted in Fig.~\ref{invert}, together with that from Paper I and the predictions of Models A, C and F of FAL93. For these models the optical depth associated to a particular height (and hence to a particular $T_e$) was computed using the opacity derived from the model parameters.

Although the present results are not identical to those of Paper I, they confirm the main conclusion that the ALMA inversion for the average QS gives lower $T_e$ than the FAL93 C model (this time $\sim240$\,K lower at 100\,GHz, compared to $\sim350$\,K reported in Paper I).

\begin{figure*}%[h]
\centering
\includegraphics[width=\textwidth]{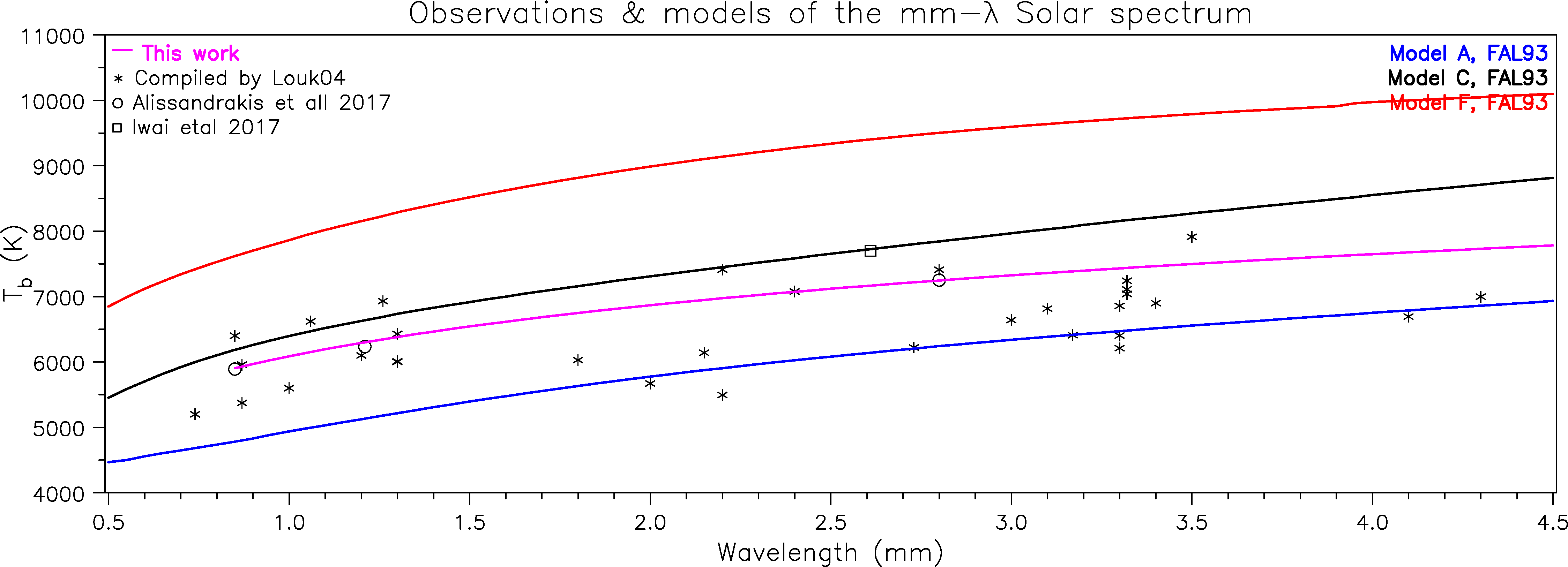}
\caption{The mm-$\lambda$ $T_b$ spectrum derived from the ALMA observations together with the prediction of the FAL 93 models and observations. Louk04 stands for \cite{2004A&A...419..747L}.}
\label{spectrum}
\end{figure*}

Furthermore, applying (\ref{reduce}) to (\ref{tbmu}), we obtain: 
\be
T_b(\mu,f)=a_1+a_2 \left\{\ln \left[\left(\frac{f}{f_{ref}}\right)^2\,\mu\right]-\gamma\right\}
\label{spec}
\ee
which, given the inversion parameters $a_1$ and $a_2$, allows the computation of the center-to-limb variation (CLV) for a given frequency, as well as the spectrum at disk center, $T_b(1,f)$. We note that (\ref{tbmu}), and hence (\ref{spec}), implicitly assumes a plane-parallel atmosphere which, according to out tests with model computations, is valid up to $\mu\simeq0.15$ ($\sim10$\arcsec\ from the limb),

The derived disk center spectrum is plotted in Fig.~\ref{spec}. The wavelength range in which this spectrum is valid is limited by our range of $\tau_{100}$ and extends from 353\,GHz to 62.5\,GHz (0.85\,mm to 4.8\,mm). In the same figure we have plotted the spectra from the FAL93 models, the set of observations compiled by \cite{2004A&A...419..747L}, the value reported by \cite{2017SoPh..292...22I} at 2.6\,mm from the 45\,m Nobeyama dish and our disk center measurements from paper~I, corrected by the normalization factors of Table~\ref{Table02}. The FAL93 $T_b$ was computed by integration of the transfer equation, using opacities derived from the tabulated values of the physical parameters.

The compilation of \cite{2004A&A...419..747L} is a highly inhomogeneous data set and has a lot of scatter. Most points fall between models A and C of FAL93; several points are near our spectrum, but most of them are below. The measurement of \cite{2017SoPh..292...22I} is above ours and the corrected Paper I values fall on our curve, as expected.

\begin{table}[!h]
\begin{center}
\caption{Disk center brightness}
\label{Table04}
\begin{tabular}{lcc}
\hline 
Band&Freq&$T_b$ \\
&GHz&K\\
\hline 
Band 3 BB1 & ~93 & 7406 \\
Band 3 BB2 & ~95 & 7382 \\
\bf{Band 3 Aver}& \bf{100} & \bf{7324} \\
Band 3 BB3 & 105 & 7269 \\
Band 3 BB4 & 107 & 7248 \\
Band 6 BB1 & 230 & 6386 \\
Band 6 BB2 & 232 & 6376 \\
\bf{Band 6 Aver}& \bf{239} & \bf{6343} \\
Band 6 BB3 & 246 & 6310 \\
Band 6 BB4 & 248 & 6301 \\
\hline 
\end{tabular}
\end{center}
\end{table}

Using (\ref{spec}) we computed disk center values of $T_b$ at some frequencies of interest for ALMA, and we give them in Table~\ref{Table04}; BB1 to BB4 refer to the 4 basebands (spectral windows) and bold characters to the average frequency of each spectral band.

The Band 3 average value (7324\,K) is very close to the recommended (7300\,K), whereas the Band 6 average value (6343\,K) is 443\,K or 7.5\% above the recommended (5900\,K). \citet[Section 9.3]{2017SoPh..292...88W} claim a rather small statistical uncertainty, but a systematic uncertainty of order 5\% comes in through the product of the ``forward scattering and spillover'' coefficient and the ``forward efficiency''. It appears that this uncertainty estimate is too low. 

\subsection{Height of the mm-$\lambda$ emission}
It is important to associate the electron temperature derived from the ALMA observations to the height, $z$, in the atmosphere. However, the computation of $\tau(z)$ requires knowledge of the absorption coefficient which, in turn, depends upon the electron, ion and H$^-$ densities that we do not have.  A direct geometric measurement of the height from the shift of the mm-$\lambda$ features with respect to the associated magnetic features, as was done by \cite{2019SoPh..294..161A} for AIA images, would be possible, were the ALMA solar pointing accurate enough. 

\begin{table*}%[!h]
\begin{center}
\caption{Estimates of emission height from the $\tau_{5000}=1$ level in the mm range and from AIA 1600 and 304\,\AA\ images}
\label{Table10}
\begin{tabular}{llcll}
\hline 
$\lambda$& Location    &Height              &   Method       & Reference \\
          &            & Mm                 &                &           \\
\hline 
1600\,\AA & Disk center & $0.4 \pm0.1$       & Network shift  & \cite{2019SoPh..294..161A} \\
          & Near limb   & $0.8\pm0.1$       & Network shift  & \cite{2019SoPh..294..161A} \\
          & Limb        & $1.4\pm0.2$       & Direct         & \cite{2019SoPh..294...96A} \\
1.26\,mm  & Limb        & $2.4\pm1.7$ & Solar radius   & This work \\
          & Limb        & $3.7\pm0.1$       & Eclipse        & \cite{1993ApJ...403..426E} \\
1.46\,mm  & Limb        & $5.3\pm2.0$       & Solar radius   & \cite{2017SoPh..292..195M}\\
3.00\,mm  & Limb        & 4.2 $\pm2.5$ & Solar radius   & This work \\
          & Limb        & $5.8\pm0.6$       & Eclipse        & \cite{1992ApJ...400..692B} \\
          & Disk        & $1.9\pm0.9$       & Oscillations   & Patsourakos et al. (2020) \\
304\,\AA\ & Disk center & $3.5\pm0.2$       & Network shift  & \cite{2019SoPh..294..161A} \\
          & Near limb   & $4.4\pm0.9$       & Network shift  & \cite{2019SoPh..294..161A} \\
          & Limb        & $5.7\pm0.2$       & Peak intensity & \cite{2019SoPh..294...96A} \\
\hline 
\end{tabular}
\end{center}
\end{table*}

Alternatively, information on the height can be provided by visual comparison of the structure in ALMA images to the structure in AIA images. In this way we estimated that the mm-$\lambda$ emission forms between the AIA levels at 1600 and 304\,\AA\ (Paper I and Paper II). An indirect estimate of the emission height can be made from the delay of oscillations observed by ALMA with respect to those observed in the AIA 1600\,\AA\ band \citep{2020A&A...634A..86P}, assuming that they are manifestations of propagating waves. Finally, measurements of the solar radius offer another estimate of the formation height, but (a) the radius reflects the maximum height of formation rather than the average, as discussed by \cite{2019SoPh..294..161A}, and (b) cannot be accurately determined from the low-resolution FD images with ALMA.

In Table~\ref{Table10} we compiled measurements relevant to the emission height in the mm range and from AIA images; the  values near or at the limb have been corrected for the $\sim340$\,km height difference between the optical limb and the $\tau_{5000}=1$ level. The average radius from the present data set was determined by fitting the gradient of each FD image with a circle (see also comments in Paper I), thus representing the position of the inflection point of the center-to-limb intensity variation. The accuracy of these values is certainly better than the resolution of the FD images; an estimate can be obtained from the rms of the deviations from the circular fit, which are $\sim1.7$\,Mm for Band 3 and $\sim2.5$\,Mm for Band 6.

The ALMA values in Table~\ref{Table10} are slightly larger than those reported in Paper I, from commissioning observations alone. These values are between the ones given by \cite{2019SoPh..294..161A} for the limb height in the 1600 and 304\,\AA\ AIA bands, confirming our assertion in Papers I and II. We note, however, that the heights given here are smaller than those reported by \cite{1993ApJ...403..426E} at 0.85\,mm and by \cite{1992ApJ...400..692B} at 3\,mm. Their measurements were obtained in eclipse observations, using single dish \citep{1993ApJ...403..426E} and interferometric methods \citep{1992ApJ...400..692B} and they refer to a particular position angle and not to the average quiet Sun. The difference between the emission heights of Bands 3 and AIA 1600\,\AA, 1.9\,Mm, of \cite{2020A&A...634A..86P} is also consistent with Band 3 forming between the 1600\,\AA\ level (0.4-0.8\,Mm on the disk) and the 304\,\AA\ level (3.5\,Mm on the disk).

\section{Cell interior and network emission}\label{HR}

\subsection{What to compare with what}
Although the comparison of the observations with models is fairly straight-forward for the average QS, the situation is more complicated in the case of cell/network measurements. A first question is what to measure. The ideal would be to use the histogram of $T_b$ values, but this is highly influenced by the resolution of the observations. The next best thing is the moments of the $T_b$ distribution, such as the rms value (related to the width of the histogram); still, this is also affected by the instrumental resolution. Probably the best choice is to use a segregation scheme among cell interior and network intensities and this is our choice for this work; it should be rather immune to instrumental effects, but then one has to decide on how to make the split. One possibility is to split the pixels equally in ``cell'' and ``network'', but it is also possible to add one or more intermediate categories. If comparison with a model is intended, the split should correspond to that of the model.

The first multi-component model was that of VAL81, who computed 6 models (Table~\ref{Table05}), based on specific ranges of pixel intensities in {\it Skylab} data (spatial resolution of $\sim5$\arcsec) in the Lyman continuum at 900\,\AA\ (see their Figure~7). We note that the Skylab resolution was inferior to that of the ALMA HR images. The models of FAL93 are based on the VAL81 segregation. Note, however, that Model~A hardly qualifies as ``quiet'' Sun, and the same thing is true for Model F which is not ``network'' (Table~\ref{Table05}). On the other hand, Model C is close to the weighed mean of the others and appears to be a good representation of the average QS (see Table~8 in VAL81).

\begin{table}%[!h]
\begin{center}
\begin{small}
\caption{The VAL 81 models}
\label{Table05}
\begin{tabular}{lccl}
\hline 
 Model &Pixels & Accumulated&Feature\\
          &\%     &    \%           &\\
\hline 
   A &  8 &    8 & Dark point within a cell\\
   B & 30 &   38 & Average cell center\\
   C & 30 &   68 & Average quiet sun \\
   D & 19 &  87 & Average network\\
   E &  9  & 96 & Bright network element\\
   F &  4  &100 & Very bright network element\\
\hline 
\end{tabular}
\end{small}
\end{center}
\end{table}

Another set of models, this time based on {\it SOHO/SUMER} observations of higher than {\it Skylab} spatial resolution, was computed by \citet[hereafter F09]{2009ApJ...707..482F}. Their model B, with index 1001, characterized as ``Quiet-sun inter-network''  comprises the 75\% lowest pixels, which is too high to qualify as inter-network QS; for example, adding the percentage of pixels attributed to dark and average cell and half of those attributed to the quiet sun by the VAL81 scheme, we get 53\%. Model D of F09, with index 1002, ``Quiet-sun network lane'', includes values from 75\% to 97\% (see their Figure~1). An additional and more serious problem with the F09 models is that they all predict a very flat chromosphere, with too low temperature gradient which, like the model of \citet[hereafter AL08]{2008ApJS..175..229A} considered in Paper I (Fig. 5), is not compatible with the ALMA measurements.

Turning to the observations, we note that time averaged images should be used, in order to avoid the influence of noise, oscillations \citep{2020A&A...634A..86P} and transient brightenings \citep{2020arXiv200407591N}. We must also bear in mind that the available ALMA results refer to a few small regions (usable FOV of about  80\arcsec\ for Band 3) at various locations, most of them in ALMA Band 3. This has an impact on the statistics, due to the small FOV; probably more important is the possibility that the locally measured values may not be representative of the average cell and network properties. Far from the disk center, we also have projection and obscuration effects.

\subsection{Processing and results}
In this work we used 10-min time-averaged images with 3-4\arcsec\ resolution from our interferometric observations of March 16, 2017, obtained in Band 3 at 6 locations (targets) on the solar disk along a position angle of 135\degr\ (Table 1 of Paper II), excluding Target 1 which was at the limb; we also used our observations of April 12, 2018, with a single target at $\mu$=0.9, at 100 and 239 GHz; these images had a resolution of $\sim2$\arcsec\ in Band 3 and $\sim1$\arcsec\ in Band 6. In both cases we used ``feathered'' images, improved by self-calibration; due to the averaging the noise level was very low. Band 3 images were measured over a circular FOV of 40\arcsec\ radius and total area of 5000\arcsec$^2$, whereas the radius used for Band 6 was 17.5\arcsec\ and the corresponding area was 960\arcsec$^2$. We preferred	 a circular, rather than a square FOV (as we had done in Paper I), in order to avoid problems due to the primary beam correction near the corners of the square. 

In order to increase the number of measurements, particularly in Band 6, we also used two mosaics near the limb, at 100 and 239 GHz, obtained during commissioning. The Band 3 mosaic, with a 190 by 178\arcsec\ FOV and a 4.6\arcsec resolution, provided usable data for $\mu\le0.55$, whereas the Band 6 mosaic, with a FOV of 142 by 75\arcsec and a 1.3\arcsec resolution, provided usable data for $\mu\le0.38$. Measurements were performed in strips of width $\Delta\mu=0.05$ for Band 3, giving projected areas from 4500 to 2700\arcsec$^2$ from $\mu=0.55$ to $\mu=0.25$. For the Band 6 mosaic we used $\Delta\mu=0.25$ and had projected areas from 1250 to 900\arcsec$^2$ from $\mu=0.375$ to $\mu=0.275$.

\begin{figure}%[!h]
\centering
\includegraphics[width=\hsize]{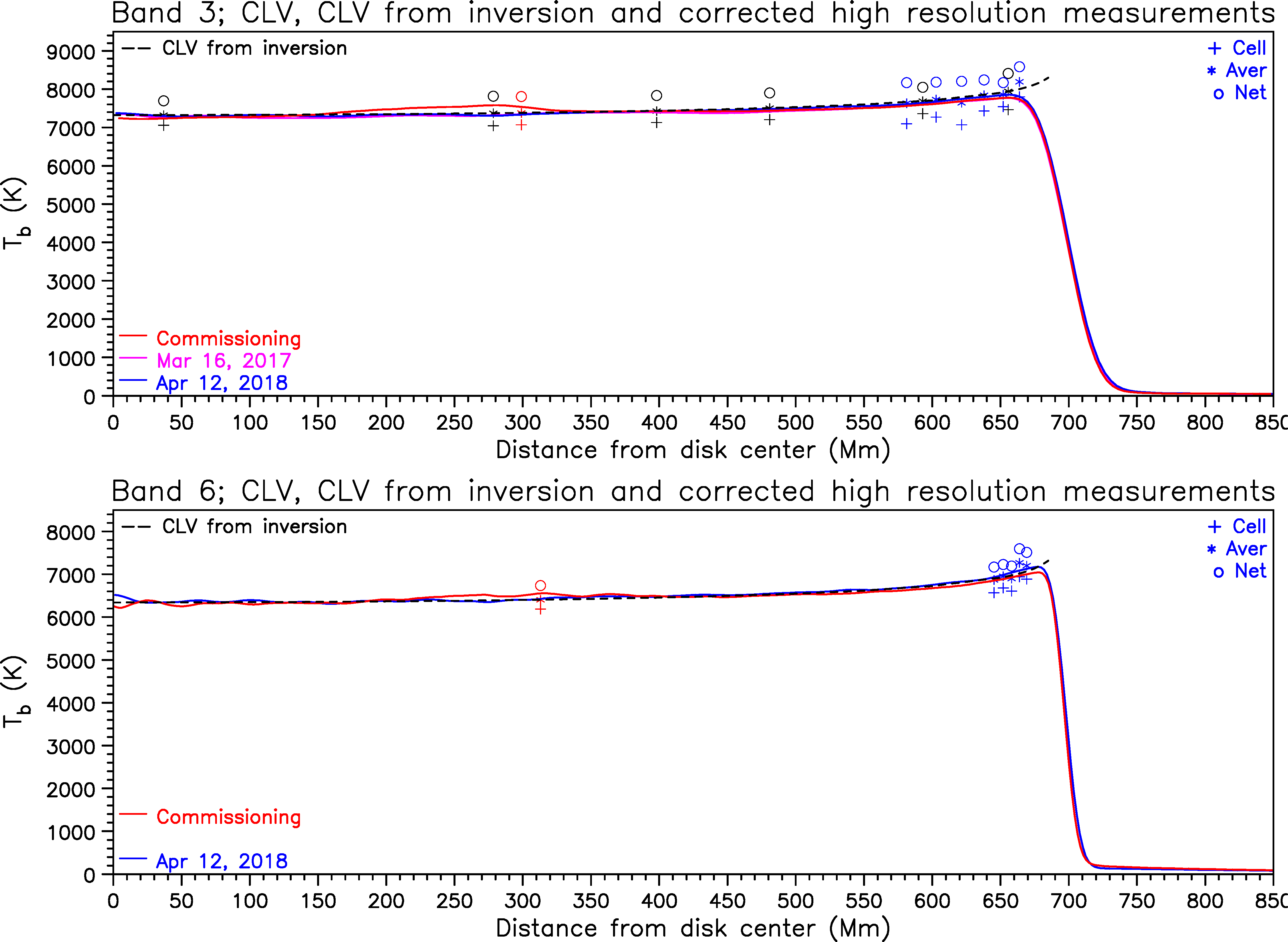}
\caption{Corrected brightness for cell/average/network for Bands 3 and 6 from HR images, as a function of distance from the center of the disk, overplotted on the corresponding CLV curves and the CLV curve derived from the inversion (dashed line). Magenta is for the March 2017 data, blue for April 2018 and red for commissioning.}
\label{CLV2}
\end{figure}

The new measurements for 2017 practically coincide with the values reported in Paper II. However, the average $T_b$ of the HR ``feathered'' images was slightly different from the corresponding value of the average center-to-limb variation (CLV) curves at the same $\mu$. We attribute this effect to differences between local conditions, reflected in the HR images, and the average QS conditions, reflected in the azimuthally averaged CLV curves. Therefore we corrected the HR values by making the average of each region equal to the prediction of the global inversion. As for the commissioning HR data, we had already pointed out in Paper I that they were above the average CLV curve; a single correction factor was used for all regions in the same band. We note that these corrections affect little the net--cell difference and not at all the net/cell ratio.

\begin{figure}[t]
\centering
\includegraphics[width=\hsize]{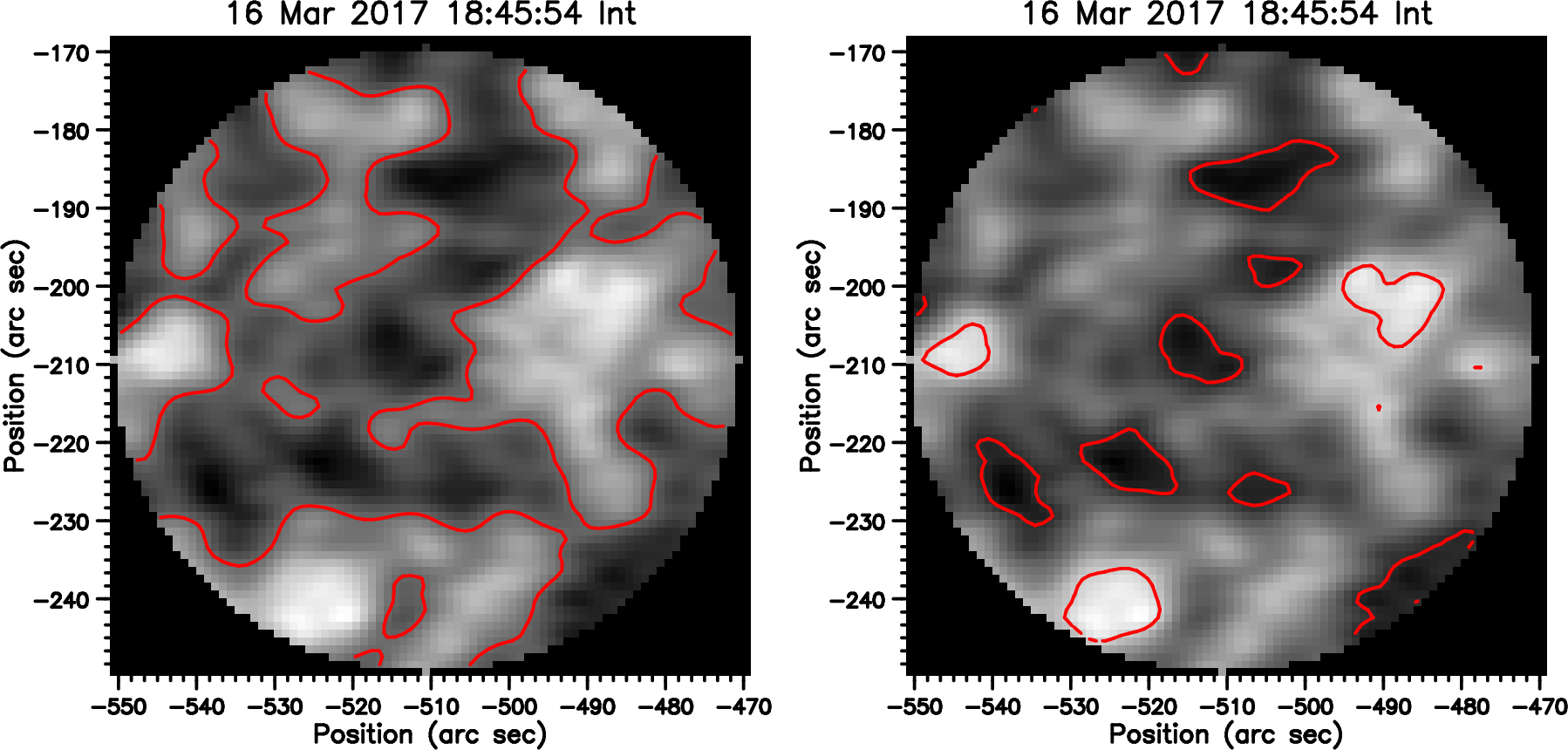}
\caption{Image of Target 5 (March 16, 2017) at 3mm, with contours corresponding to the segregation used for measurement set 1 (left) and set 2 (right). The radius of the FOV is 40\arcsec.}
\label{Segr}
\end{figure}

The corrected cell, average and network $T_b$ values are plotted in Fig.~\ref{CLV2}. In the same figure we plotted the associated CLV curves, as well as the CLV curve deduced from our inversion (dashed lines); thus this figure also serves as a check of our inversion against the actual CLVs.

Two segregation schemes were used. The first (measurement set 1) attributed equal numbers of pixels to cell interior and network (as in Paper II), whereas the second (measurement set 2) followed VAL81 Models A and F by attributing the lowest 8\% of the pixels to cell interior and the top 4\% to the network; this choice is justified by the fact that, as pointed out in Paper II, the mm-$\lambda$ structure is very similar to that in the UV continuum. Set 2 gives worse statistics than set 1, but it was necessary in order to come as close as possible to the FAL93 models. Fig.~\ref{Segr} gives contours of the cell-network boundaries for both segregation schemes, superposed on a Band 3 image. Although the second scheme might appear extreme, it picks up well the darkest cell and the brightest network pixels, but not the average cell and network. The measurements are tabulated in Table~\ref{Table06}, where some values from set 2 with excessive deviation have been deleted.

\begin{table*}
\begin{center}
\begin{small}
\caption{Brightness temperature for cell interior and network from high-resolution observations, $\mu_{100}$ is the value of $\mu$ reduced to 100\,GHz.}
\label{Table06}
\begin{tabular}{lccccccl}
\hline 
    &     &         &\multicolumn{2}{c}{Set 1}&\multicolumn{2}{c}{Set 2}\\
Freq&$\mu$&$\mu_{100}$&$T_b$ Cell&$T_b$ Net&$T_b$ Cell&$T_b$ Net&Origin\\
GHz &       &      &   K   &  K     &   K    &  K     &                            \\
\hline
100 & 0.250 & 0.25 & 7991 &  8871 &  7048 &  9296 & Commissioning Band 3 mosaic\\
    & 0.300 & 0.30 & 7796 &  8584 &  7354 &  9030 &                            \\
    & 0.350 & 0.35 & 7544 &  8168 &  7172 &  8613 &                            \\
    & 0.400 & 0.40 & 7432 &  8239 &  6907 &  8692 &                            \\
    & 0.450 & 0.45 & 7072 &  8203 &   --  &  8733 &                            \\
    & 0.500 & 0.50 & 7270 &  8182 &  6827 &  8877 &                            \\
    & 0.550 & 0.55 & 7095 &  8169 &  6681 &  8901 &                            \\
    & 0.340 & 0.34 & 7464 &  8411 &   --  &   --  &  March 16, 2017 (Paper II)  \\
    & 0.520 & 0.52 & 7361 &  8046 &  6953 &  8607 &                            \\
    & 0.720 & 0.72 & 7206 &  7902 &  6801 &  8557 &                            \\
    & 0.820 & 0.82 & 7128 &  7833 &  6772 &  8504 &                            \\
    & 0.920 & 0.92 & 7043 &  7817 &  6635 &  8541 &                            \\
    & 1.000 & 1.00 & 7056 &  7694 &  6734 &  8273 &                            \\
    & 0.900 & 0.90 & 7071 &  7809 &  6704 &  8564 & April 12, 2018             \\
239 & 0.275 & 1.57 & 6595 &  7439 &  6460 &  7988 & Commissioning Band 6 mosaic\\
    & 0.300 & 1.71 & 6889 &  7511 &  6449 &  8020 &                            \\
    & 0.325 & 1.86 & 6942 &  7595 &  6314 &  7706 &                            \\
    & 0.350 & 2.00 & 6609 &  7193 &  6377 &  7668 &                            \\
    & 0.375 & 2.14 & 6680 &  7229 &  6182 &  7663 &                            \\
    & 0.890 & 5.27 & 6187 &  6738 &  5934 &  7352 & April 12, 2018             \\
\hline
100 & 0.980 & 0.98 & 6440 &  7853 &       &       & Derived from \cite{2019ApJ...877L..26L}     \\
      & 1.000 & 1.00 & 7228 &  7558 &       &       & Wedemeyer et al. (2020)  \\
\hline
\end{tabular}
\end{small}
\end{center}
\end{table*}

\begin{figure}%[!h]
\centering
\includegraphics[width=\hsize]{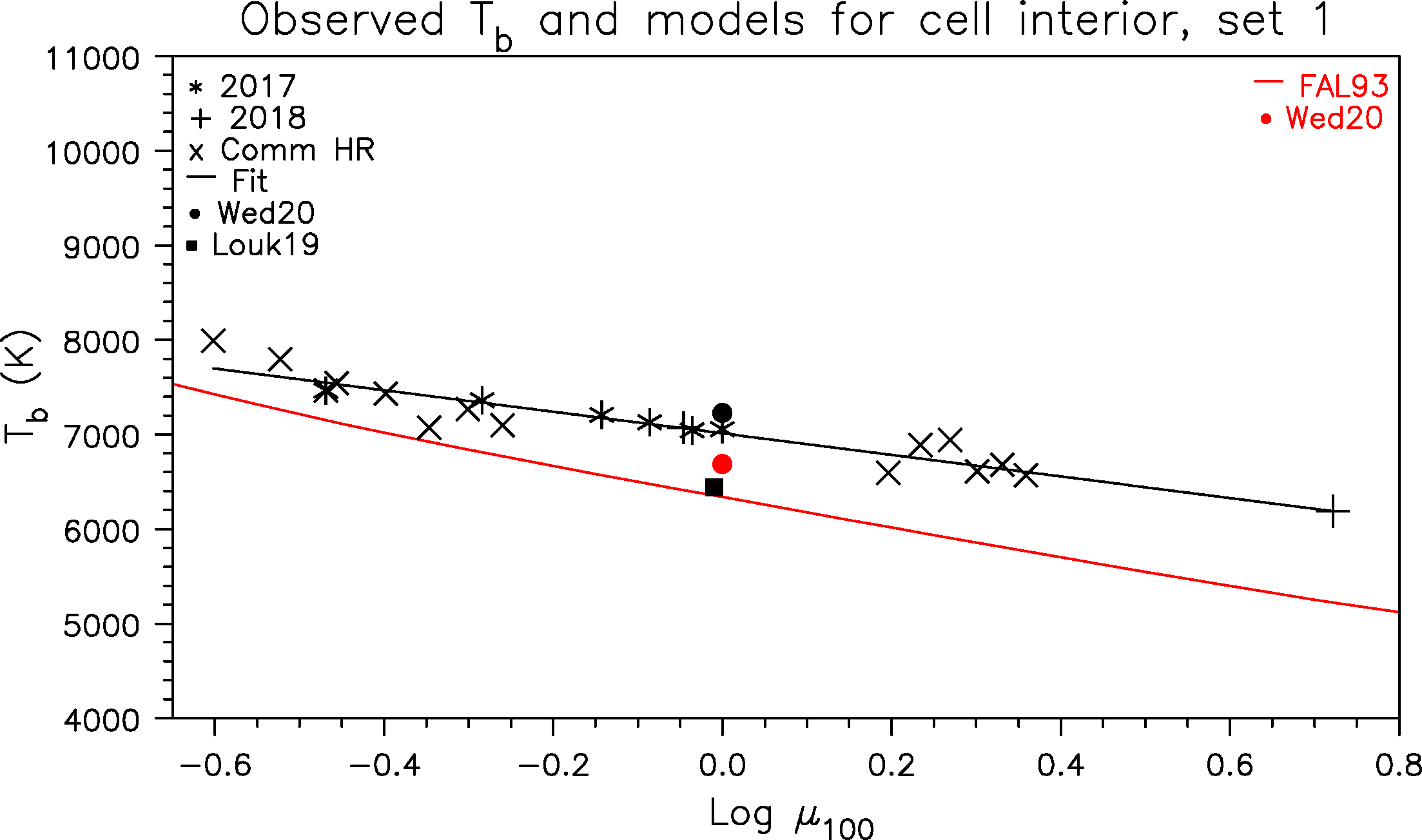}
\includegraphics[width=\hsize]{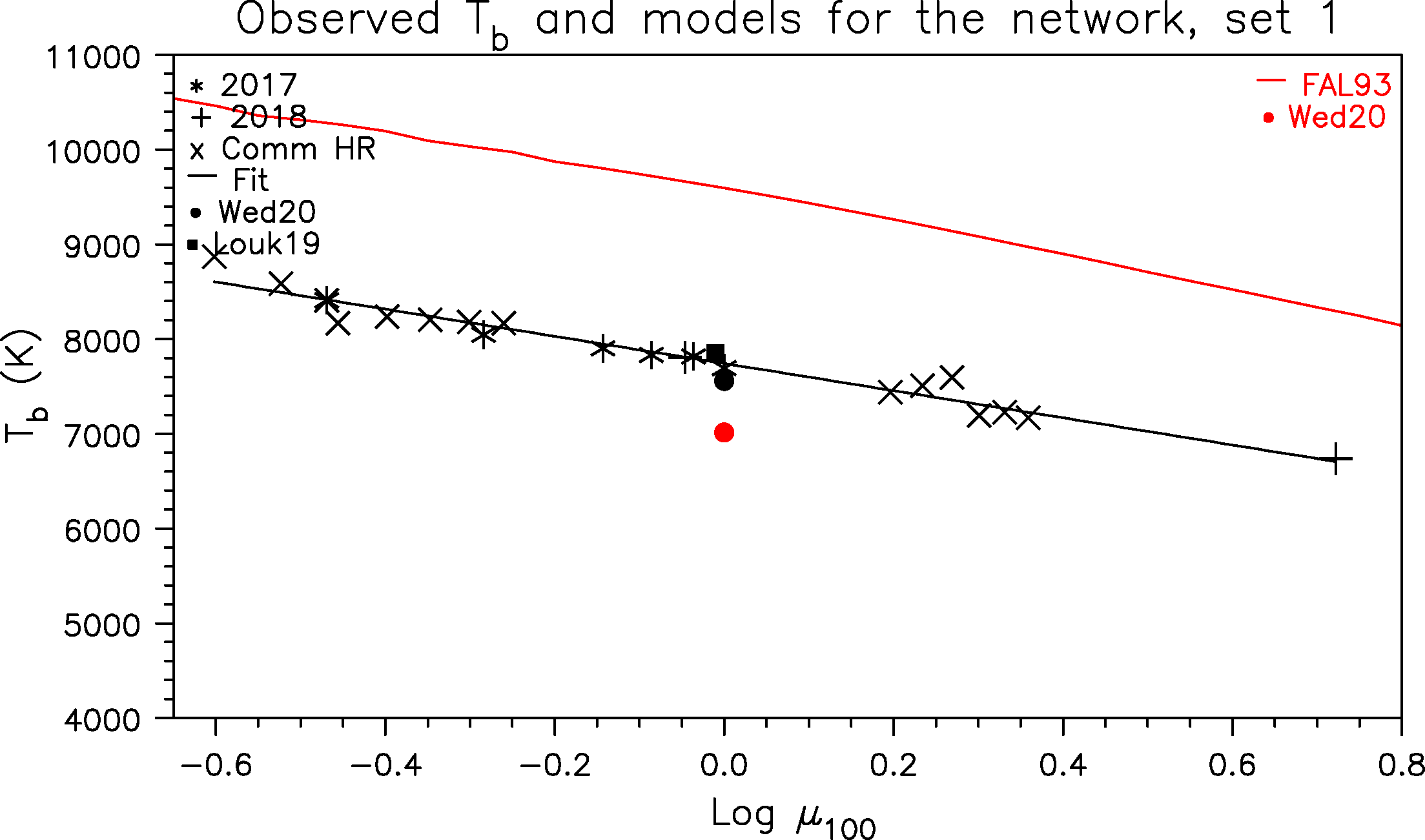}
\caption{Brightness temperature as a function of reference $\mu$ for the cell interior (top) and for the network (bottom) for measurement set 1.  Different symbols denote different data sets, from commissioning mosaics, from March 16, 2017 and from April 12, 2018. The black line is the result of linear regression. Values from \citet[][Louk19]{2019ApJ...877L..26L} and from \citet[][Wed20]{2020A&A...635A..71W} are also plotted for reference. The red full lines show the FAL93 models A (top) and F (bottom). The red filled circle is from the rMHD model of \cite{2020A&A...635A..71W}. }
\label{ResultsA}
\end{figure}

\begin{figure}[!h]
\centering
\includegraphics[width=\hsize]{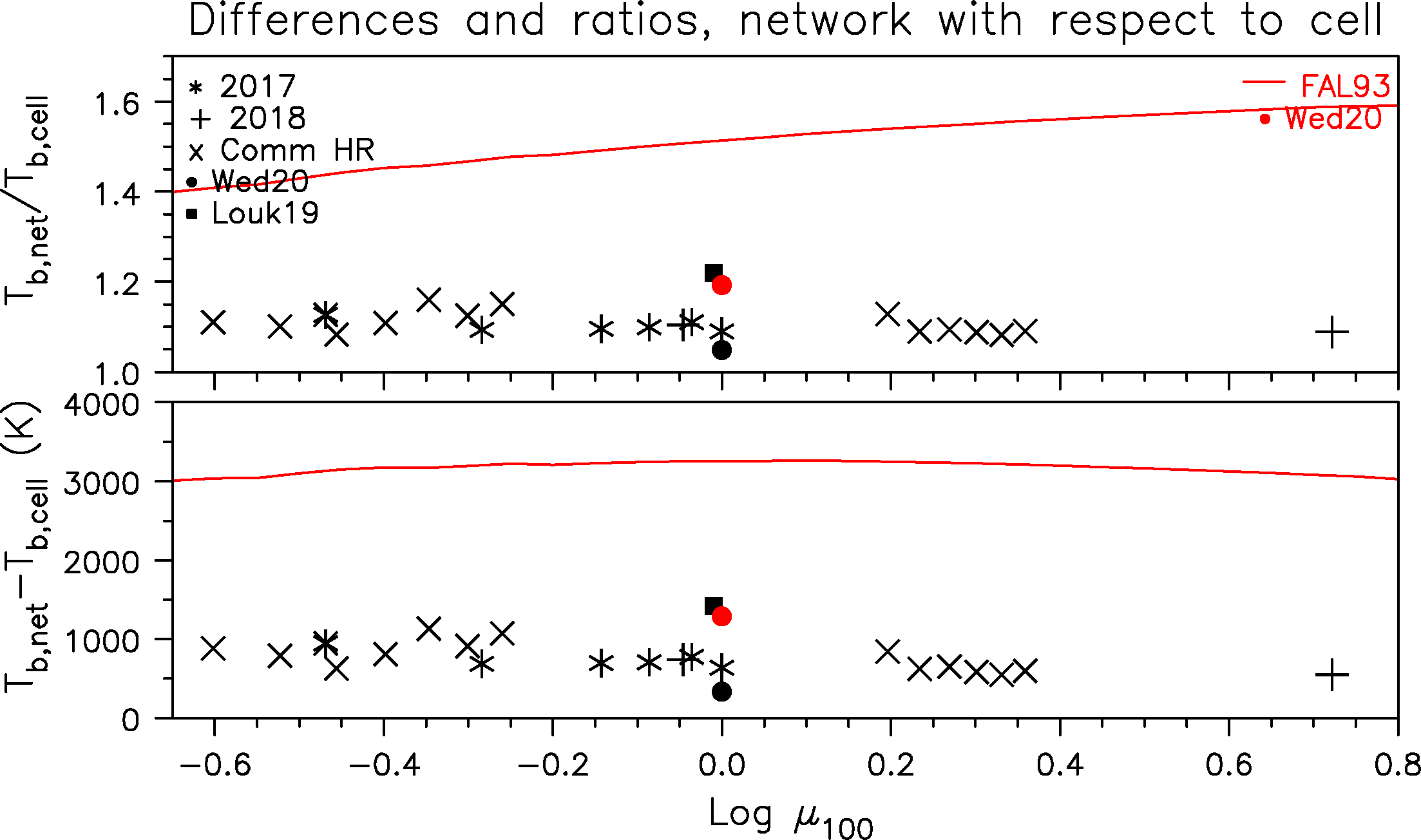}
\includegraphics[width=\hsize]{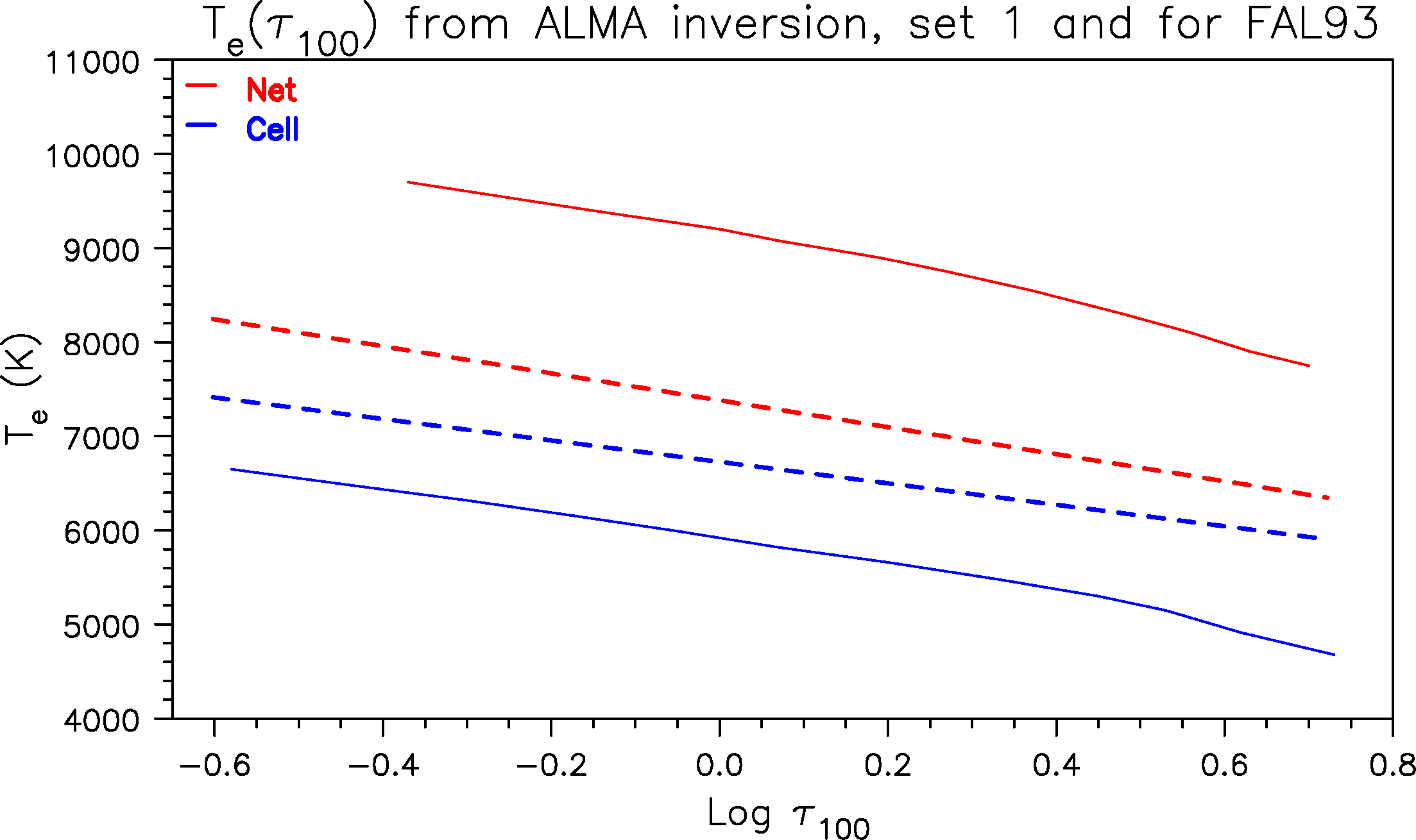}
\caption{{Top: $T_b$ differences and ratios between network and cell as a function of reference $\mu$. The symbols have the same meaning as in Fig.~\ref{ResultsA}. The red full lines show the predictions of the FAL93 models. Bottom: Thick dashed lines give the electron temperature as a function of reference $\tau$ from the inversion of the observations for the cell interior (blue) and for the network (red); thin full lines show the FAL93 models A and F.
\vspace{-.4cm} }}
\label{ResultsB}
\end{figure}

Plots of the measurement set 1 for the cell interior and the network are given in Fig.~\ref{ResultsA}, together with the corresponding linear regression lines and FAL93 models. For reference, we also plotted in the same figure and listed in Table~\ref{Table06} values derived from \cite{2019ApJ...877L..26L} at 100\,GHz near the center of the disk, measured from the histograms of their Figure~2a; we considered their ``dark region'' (which actually is the interior of a large supergranule) as cell interior, and their ``bright network'' as network. Also plotted in Fig.~\ref{ResultsA} are cell and network values reported by \cite{2020A&A...635A..71W}, observed at 100\,GHz at the center of the disk, using a two-level segregation scheme. Finally, in addition to the FAL93 models, the cell and network values (a single set at 100 GHz, $\mu=1$) form the radiative MHD model of \cite{2020A&A...635A..71W} are plotted; for this we did not take into account the degradation by the ALMA beam because the effect is small.

In spite of the inhomogeneity of the data set, the $T_b$ - $\log \mu_{100}$ plots of  Fig.~\ref{ResultsA} show well-defined linear relations, similar to the one we found from the full-disk data. It is also clear that the net/cell contrast (ratio of $\simeq1.1$) is much below the one predicted by the FAL93 models (ratio $>1.4$), as shown in the $T_b$ difference and ratio plots of Fig.~\ref{ResultsB}, top.  The brightness difference translates to an electron temperature difference, quite prominent in the $T_e(\tau_{100})$ curves obtained from the inversion of $T_b(\mu_{100})$ and plotted in the bottom panel of Fig.~\ref{ResultsB}. 

Comparing our measurements to the values derived from \cite{2019ApJ...877L..26L}, the ``cell'' $T_b$ is clearly but not dramatically below ours (expected because this is a peculiar case) and the ``net'' value very close to ours. The values of net and cell of  \cite{2020A&A...635A..71W} are much closer to the average than ours and the contrast quite smaller than ours. Finally, the rMHD model of \cite{2020A&A...635A..71W} predicts lower cell brightness and quite low network brightness compared to our measurements; it also predicts a rather high net/cell ratio of 1.19; still differences are smaller than with the FAL93 models.

\begin{figure}%[!h]
\centering
\includegraphics[width=\hsize]{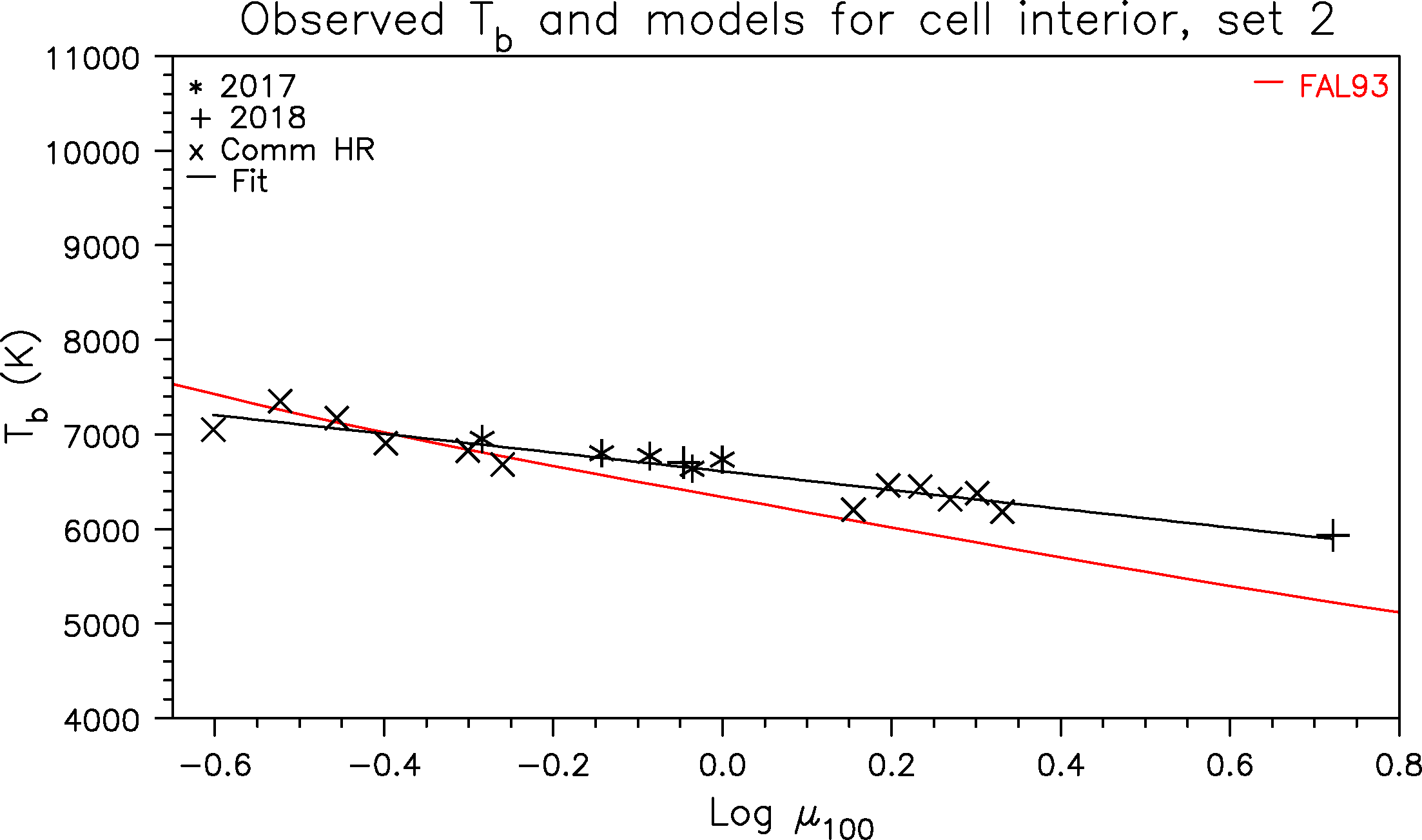}
\includegraphics[width=\hsize]{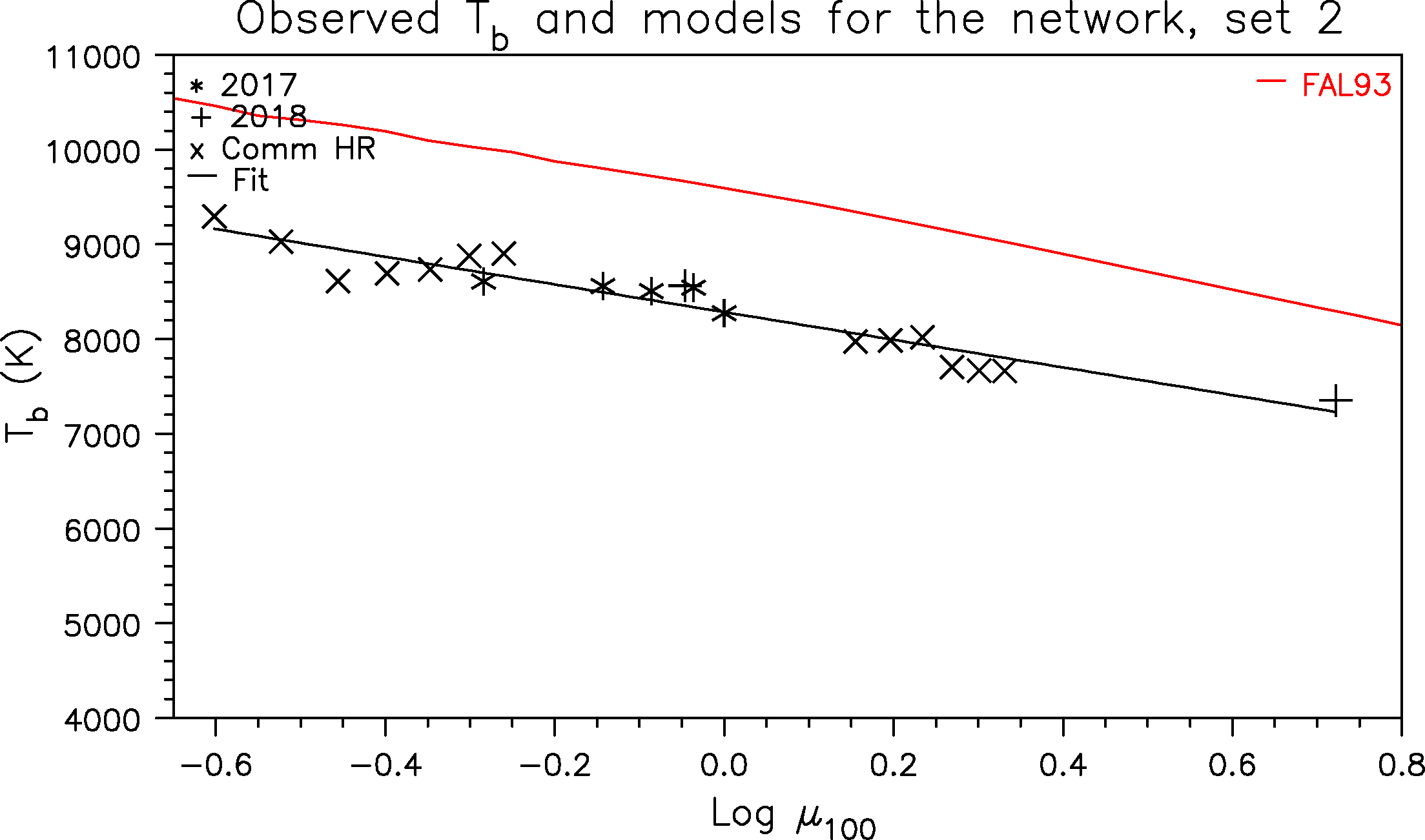}
\caption{{Same as Fig.~\ref{ResultsA} for measurement set 2.}}
\label{ResultsC}
\end{figure}

\begin{figure}%[!h]
\centering
\includegraphics[width=\hsize]{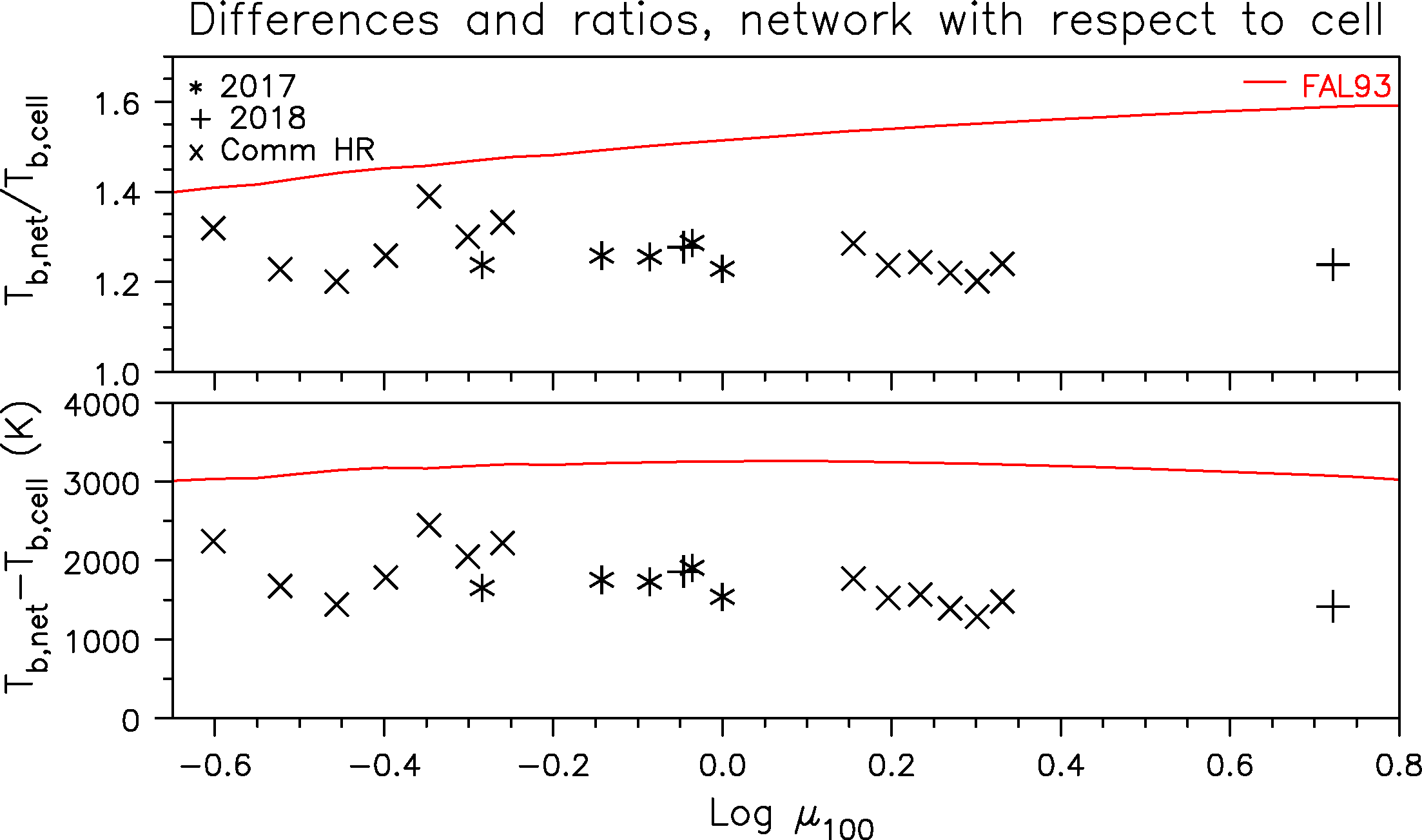}
\includegraphics[width=\hsize]{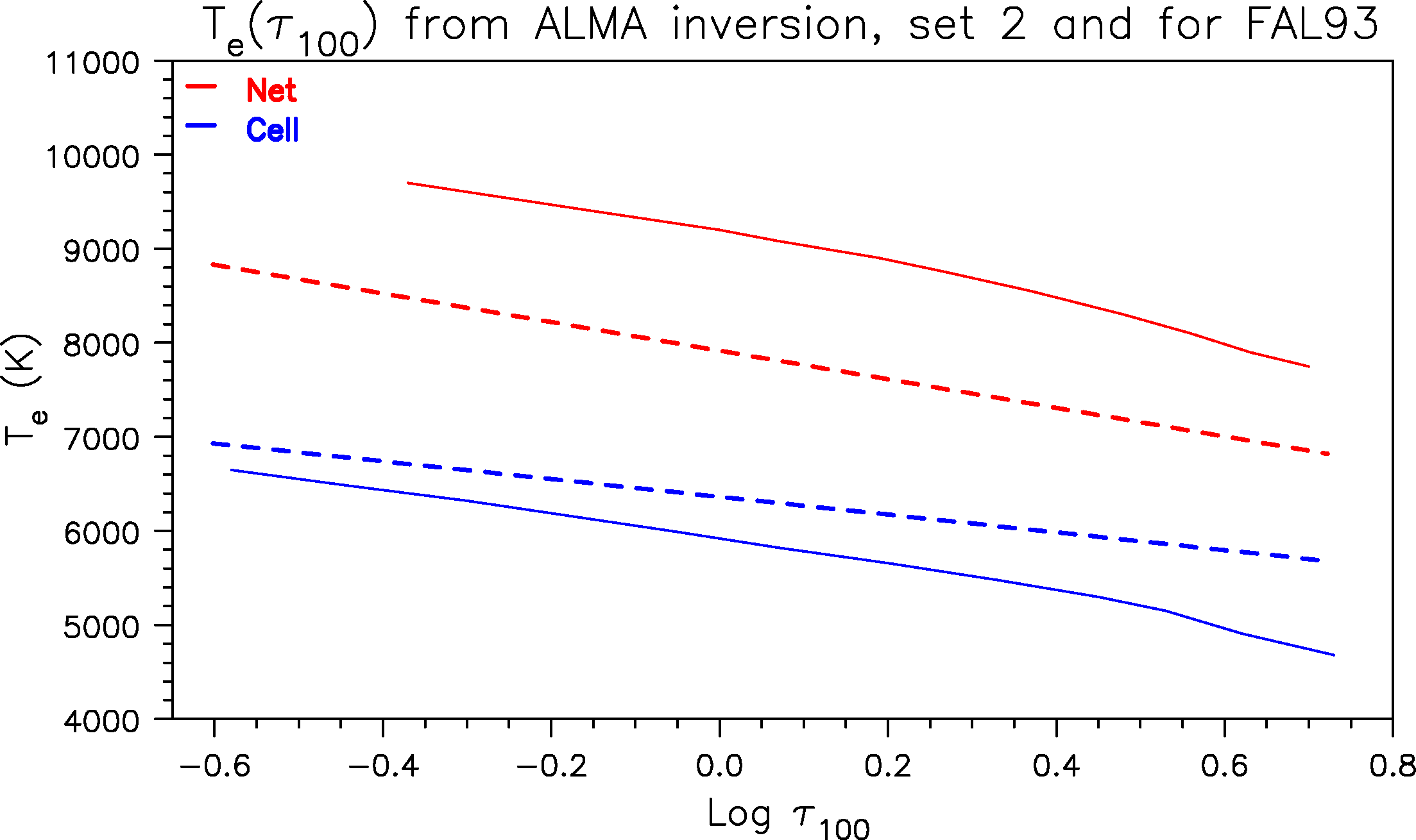}
\caption{{Same as Fig.~\ref{ResultsB} for measurement set 2.}}
\label{ResultsD}
\end{figure}

Part of the difference between our measurements and the FAL93 models could be due to the fact that, as discussed above, their observational basis does not really reflect the cell and network conditions. It is for this reason that we performed our second set of measurements, which follows the original VAL81 segregation, at the expense of inferior statistics. The results are shown in Figs. \ref{ResultsC} and \ref{ResultsD}. As expected, the cell interior in now fainter than before and the network brighter; however, the cell brightness is still above that of Model A and the network brightness still below that of model F. Although the net/cell contrast increased above 1.2, it is still too low compared to the model prediction (Fig.~\ref{ResultsD}, top) and the inversion curves, the network curve in particular, are still far from the models (Fig.~\ref{ResultsD}, bottom).

An additional remark is that for both sets the cell-network inversion curves diverge as $\tau$ decreases, indicating that the contrast decreases with $\tau$ and increases with wavelength, following the trend in the microwave range (Sect.~\ref{intro}). This may also indicate an increase of the contrast with height (note though that equal $\tau$ does not necessarily imply equal height). Associating the slope of the $T_e(\tau_{100})$ curve with the temperature gradient in the chromosphere, this result implies that the temperature rise is steeper in the network than in the cell interior.

The values of the inversion parameters $a_1$ and $a_2$ in (\ref{reduce}), which correspond to the value of $T_e$ at $\tau_{100}=1$ and $dT_e/d\ln\tau$ respectively, are given in Table~\ref{Table07}, together with the corresponding parameters of the FAL93 and AL08 models; for the models, the temperature gradient at $\tau_{100}=1$ is also given. We estimate that the errors in $a_1$ and $a_2$ are of the same order as those for the average QS, discussed in Sect.~\ref{inversion}. We note that the values of both $a_1$ and $a_2$ increase as we go from the cell interior to the network, the increase of $a_2$ reflecting the divergence between the network and cell $T_e(\tau)$ curves mentioned above.

The differences in $T_e$ between the ALMA results and the FAL93 models together with the corresponding ratios are given in Table~\ref{Table08}, while those between the network and cell interior are presented in Table~\ref{Table09}. Both tables show very well the departure of our measurements from the A and F model predictions.

\begin{table}%[!h]
\begin{center}
\caption{Atmospheric parameters from ALMA inversion and models}
\label{Table07}
\begin{tabular}{lccc}
\hline 
Model/observed & $T_e(\tau_{100}=1)$ & $dT_e/d\ln\tau$&$dT_e/dz$ \\
                  &          K          &        K    &   K\,km$^{-1}$\\
\hline 
Network, set 2 & 7916 & $-660$ &  --  \\
Network, set 1 & 7383 & $-580$ &  --  \\
Average QS     & 6999 & $-563$ &  --  \\
Cell, set 1    & 6726 & $-495$ &  --  \\
Cell, set 2    & 6362 & $-410$ &  --  \\
\hline 
FAL93, F        & 9200 & $-642$ & 1.64 \\
FAL93, C        & 7241 & $-560$ & 2.10 \\
FAL93, A       & 5918 & $-579$ & 9.11 \\
\hline 
AL08           & 6657 & ~$-15$ & 0.11 \\
\hline 
\end{tabular}
\end{center}
\end{table}

\begin{table}%[h]
\begin{center}
\caption{$T_e$ differences and ratios at $\tau_{100}=1$ with respect to FAL93}
\label{Table08}
\begin{tabular}{lcc}
\hline 
Data set       & Difference & Ratio \\
               &     K       &        \\
\hline 
Network, set 2 & $-1284$ & 0.86 \\
Network, set 1 & $-1817$ & 0.80 \\
Average QS     & ~$-242$ & 0.94 \\
Cell, set 1    &  ~~808  & 1.34 \\
Cell, set 2    &  ~~444  & 1.08 \\
\hline 
\end{tabular}
\end{center}
\end{table}

\begin{table}%[h]
\begin{center}
\caption{Network/cell interior $T_e$ differences and ratios at $\tau_{100}=1$}
\label{Table09}
\begin{tabular}{lcc}
\hline 
Data set/model  & Difference & Ratio \\
               &     K       &        \\
\hline 
set 2          &   1554  & 1.24 \\
set 1          &    657  & 1.10 \\
FAL93          &   3282  & 1.55 \\
\hline 
\end{tabular}
\end{center}
\end{table}

\section{Summary and discussion}\label{summary}
Using a larger data set of ALMA full-disk images than in Paper I, we verified that the brightness temperature varies linearly with the logarithm of $\mu$, reduced to the reference frequency of 100\,GHz, over a range of $\mu_{100}$ between $\sim0.4$ and $\sim6.6$. This implies a linear relationship between the electron temperature and the logarithm of the optical depth in the region of formation of the radiation (chromosphere).

Further study of the FD images revealed that the normalization factor recommended by \cite{2017SoPh..292...88W}
for ALMA Band 6 is underestimated by 7.5\%, or $\sim440$\,K. This conclusion was reached by fitting all measurements to the same linear $T_b$ - $\log\mu$ relation, assuming that the commissioning calibration was correct. This assumption introduces some uncertainty in our results which, however, requires a better absolute calibration of ALMA FD images in order to be avoided.

The inversion of all our FD data confirmed our original conclusion that the electron temperature in the chromosphere is close to the prediction of Model C of FAL93 (240\,K below at 100\,GHz, or 3\%); moreover, the slope of the $T_e(\tau_{100})$ curve is incompatible with other models, such as those of VAL81, AL08 and F09, which all predict too flat a chromosphere. The similarity of the present results with model C, allows us to assert that the average temperature gradient in the chromosphere is of the same order as the one predicted by that model, $\sim2$\,K\,km$^{-1}$.

From our inversion parameters we computed the mm-$\lambda$ $T_b$ spectrum at the center of the solar disk and compared it with published measurements. We found that several observed values are close to this spectrum, although most are below.

The solar radius measured from our present full-disk data set gave improved values of the limb height compared to Paper I, 2.4\,Mm in Band 6 and 4.2\,Mm in Band 3; this confirms our assertion in Papers I and II that the mm-emission forms between the levels of the 1600\,\AA\ and 304\,\AA\ emissions. Still, more work is necessary in this direction, because of the low resolution of the ALMA FD images.

Measurements of the intensity of cell interiors and network elements from high resolution ALMA data in bands 3 and 6 revealed similar linear relationships between $T_e$ and $\log(\tau_{100})$, with lower slope for the cell interior and higher for the network, compared to the average QS. The divergence of the  $T_e(\log \tau_{100})$ curves indicates that the temperature rises faster in the network than in the cell interiors. 

Our measurements give a much lower network-to-cell contrast than the FAL93 models, even if we use the pixel segregation scheme of VAL81, on which the FAL93 models are based. The cell interior is brighter than the model prediction and the network less bright. This may reflect issues related to the observational basis of the models.

In addition to the new information provided in this work about the temperature structure of the chromosphere, a number of important observational issues is raised. One is that a better absolute calibration is needed for ALMA FD images. This could be achieved  through cross-calibration with the full Moon; alternatively, a comparison of FD data with HR data, which are calibrated using celestial sources, could be tried. A second issue is that statistics of HR observations should be improved, by more observations at different disk locations and in all available bands; the ideal, of course would be to obtain HR images of the entire solar disk, but this is not currently possible with ALMA.

In spite of the poor statistics, the main results of the present work, {\it i.e.} the linear dependence of $T_e$ on $\tau$, the increase of the slope as we go from the cell interior to the network, the network/cell contrast, appear robust. Expanding ALMA observations to higher frequencies (Band 7 is already in operation), will allow us to better probe the low chromospheric levels and to approach, or even reach the temperature minimum. Expanding to lower frequencies will provide information about the upper chromosphere and its interface with the transition region.

The approach that we used in Paper I and developed further here, gives physical information from CLV-spectral observations in a direct and simple way. We have not yet treated the important issue of pole-equator differences, polar brightening in particular \citep[see Sect. 4.2 in ][]{2011SoPh..273..309S} and, of course, solar cycle variations are out of ALMA reach for the time being. Finally, we expect to investigate further in the near future both the average quiet Sun and the cell/network properties at higher frequencies, as Band 7 observations become available. Band 5 data, which will become available in the upcoming Cycle 8 of ALMA solar observations, will also be very useful to bridge the gap between Bands 3 and 6.

\begin{acknowledgements}
This paper makes use of the following ALMA data: ADS/JAO.ALMA\#2011.0.00020.SV., ADS/JAO.ALMA\#2016.1.00572.S, ADS/JAO.ALMA\#2017.1.00653.S, ADS/JAO.ALMA\#2017.1.00870.S, ADS/JAO.ALMA\#2018.1.01763.S and ADS/JAO.ALMA\#2019.1.01532.S. ALMA is a partnership of ESO (representing its member states), NSF (USA), and NINS (Japan), together with NRC (Canada) and NSC and ASIAA (Taiwan), and KASI (Republic of Korea), in cooperation with the Republic of Chile. The Joint ALMA Observatory is operated by ESO, AUI/NRAO, and NAOJ.
\end{acknowledgements}

\end{document}